\documentclass[12pt]{elsart} 
 
\usepackage{amssymb} 
\usepackage{latexsym} 
\usepackage{amsfonts} 
\usepackage{graphicx} 
\usepackage{psfrag}

\newcommand{\be}{\begin{equation}} 
\newcommand{\ee}{\end{equation}} 
 
\newtheorem{theorem}{Theorem} 
 
\newcommand{\A}{\mathcal{A}} 
\newcommand{\Aia}{A^{i}_{a}}

\newcommand{\g}[1][]{\gamma_{#1}} 
 
\newcommand{\W}{\mathcal{W}}

\newcommand{\QR}{\Q|_R}

\newcommand{\Q}{\mathcal{Q}} 
\newcommand{\gq}{\Gamma_\Q}

\newcommand{\Al}{\mathfrak{A}} 
\newcommand{\I}{\mathfrak{I}} 
 
\newcommand{\Aob}{\mathfrak{B}_\mathrm{aux}} 
 
\newcommand{\alk}{\mathfrak{B}_\mathrm{kin}} 
\newcommand{\cyl}{\mathfrak{C}} 
\newcommand{\ocyl}{\overline{\cyl}} 

\newcommand{\ba}{\mathbf{a}}
\newcommand{\bb}{\mathbf{b}}

\newcommand{\Ho}{\mathcal{H}_{\omega}} 
\newcommand{\Ha}{\mathcal{H}_\mathrm{aux}} 
 
\newcommand{\Hg}{\mathcal{H}_{\G}} 
\newcommand{\Hk}{\mathcal{H}_\mathrm{kin}} 
 
\newcommand{\Hq}{\mathcal{H}_{\Q}} 
 
\newcommand{\Hqk}{\Hq^\mathrm{kin}} 
 
\newcommand{\G}{\mathcal{G}} 
\newcommand{\D}{\mathcal{D}} 
\newcommand{\Dg}{{\rm Diff}_{\Gamma}} 
 
\newcommand{\ip}[2]{\left\langle #1 | #2 \right\rangle} 
\newcommand{\norm}[1]{\left\| #1 \right\|} 
\newcommand{\snorm}[1]{\left\| #1 \right\|_{\infty}} 
 
\newcommand{\om}{\omega}

\newcommand{\po}{\pi_{\om}} 
 
\newcommand{\ra}{\rightarrow} 
 
\newcommand{\area}{\langle a \rangle} 
\newcommand{\vol}{\langle v \rangle} 
\newcommand{\Vol}{\langle V_R \rangle} 
\newcommand{\Area}{\langle A_S \rangle} 
\newcommand{\oArea}{\overline{\Area}} 
\newcommand{\oVol}{\overline{\Vol}} 
 
\newcommand{\Adev}{\sigma_A} 
\newcommand{\Vdev}{\sigma_V} 
\newcommand{\adev}{\Delta_a} 
\newcommand{\vdev}{\Delta_v} 
 
\begin{document} 
\begin{frontmatter} 
 
\title{Loop Quantum Gravity on Non-Compact Spaces} 
\author{Matthias Arnsdorf\thanksref{mat}}\\ 
\address{Blackett Laboratory, Imperial College of 
 Science Technology and Medicine,\\ London SW7 
 2BZ, United Kingdom.} 
\thanks[mat]{m.arnsdorf@ic.ac.uk} 
 
\author{Sameer Gupta\thanksref{sam}} 
\address{Center 
 for Gravitational Physics and Geometry, Physics Department \\ 
 The Pennsylvania State University, University 
 Park, PA 16802, USA.} 
\thanks[sam]{gupta@gravity.phys.psu.edu} 
 
\begin{abstract} 
We present a general procedure for constructing new Hilbert spaces for 
loop quantum gravity on non-compact spatial manifolds. Given any fixed 
background state representing a non-compact spatial geometry, we use 
the Gel'fand-Naimark-Segal construction to obtain a representation of 
the algebra of observables. The resulting Hilbert space can be 
interpreted as describing fluctuation of compact support around this 
background state. We also give an example of a state which 
approximates classical flat space and can be used as a background 
state for our construction. 
 
\end{abstract} 
\end{frontmatter} 
 
\section{Introduction} 
 
Remarkable progress has been made in the field of non-perturbative (loop) 
quantum gravity in the last decade or so and it is now a rigorously defined 
kinematical theory. One of the most important results in this area is that 
geometric operators such as area and volume have discrete spectra. However, 
before loop quantum gravity can be considered a complete theory of quantum 
gravity, we must show that the discrete picture of geometry that it provides 
us reduces to the familiar smooth classical geometry in some appropriate 
limit.  One aspect of this is the recovery of the weak-field limit of quantum 
gravity which is described by gravitons and their interactions. 
 
In standard perturbative quantum field theory, gravitons are fields which 
describe the fluctuations of the metric field around some classical ``vacuum'' 
metric (usually Minkowski space). The graviton is then a spin-two particle as 
defined by the representations of the Poincar\'e group at infinity. Thus, in 
order to study graviton physics in the context of loop quantum gravity, a 
basic requirement is the construction of a state corresponding to the 
Minkowski metric, which in turn necessitates a proper quantum treatment of 
asymptotically flat spaces. Given this framework one could then construct 
asymptotic states (corresponding to gravitons) describing fluctuations of the 
background metric and the action of the generators of the Poincar\'{e} group 
at infinity. 
 
These are non-trivial requirements in loop quantum gravity, which can
be thought of as describing excitations of the three-geometry
itself. Hence, the `zero excitation state' of the theory corresponds
to the metric ``$g_{ab} = 0$'' and not the Minkowski metric. In this
description, Minkowski space-time is a highly excited state of the
quantum geometry containing a infinite number of elementary
excitations. The situation, in fact, is analogous to that in
finite-temperature field theory where the thermal ground state is a
highly excited state of the zero-temperature theory which does not
even lie in the standard Fock space.  Excitations are then constructed
by building a representation of the standard algebra of creation and
annihilation operators on the thermal vacuum. In this paper, we shall
give an analogous construction in loop quantum gravity which enables
us to describe fluctuations of essentially compact support around a
flat background metric. It is important to note that these
fluctuations can be arbitrarily large and hence we are not quantising
linearised general relativity. Indeed, our framework aims to identify
quantum linearised general relativity as a sector of the
non-perturbative theory.
 
This work is comprised of two main parts. After a brief review of
concepts from loop quantum gravity, we describe how new Hilbert spaces
for loop quantum gravity applicable to non-compact spaces can be
constructed by finding new representations of the standard algebra of
observables given some notion of `vacuum' or `background' state.

We proceed to present a detailed example of a background state $\Q$, which
approximates the Euclidean metric on three-space.  This state is
related to
the weave construction, but differs from it as it is peaked not
only in the spin-network basis but also in the connection basis. We
show that even though this state is not an element of the standard
Hilbert space, as is generically the case for states approximating
geometries on non-compact spaces, it can be used as a ``vacuum'' in
the above construction to give genuine Hilbert spaces describing
fluctuations around this state.

\section{The Structure of Loop Quantum Gravity} 
 
Canonical general relativity can be written as a theory of a real SU(2) 
connection over an oriented three-manifold $\Sigma$ 
\cite{ashtekar86,barbero94}. The classical configuration space $\A$ is given 
by all smooth connections $A$ on a principle SU(2)-bundle\footnote{This bundle 
arises as the double cover of the principle $SO(3)$-bundle of frames on 
$\Sigma$.} $P$ over $\Sigma$. Since $P$ is trivial, we can use a global cross 
section to pull back connections to $\mathfrak{su}(2)$-valued 
one-forms\footnote{Indices $i$,$j$,$k$,\ldots denote an internal 
$\mathfrak{su}(2)$ indices, while $a$,$b$,$c$,\ldots are tensor indices.} 
$A^{i}_{a}$ on $\Sigma$.  The conjugate variable to the connection is a 
densitized triad $\tilde{E}_i^b$ which takes values in the lie algebra 
$\mathfrak{su}(2)$. The triads can be considered as duals to two-forms 
$e_{abi}~\equiv~\eta_{abc}\tilde{E}_i^c$. The dynamics of general relativity 
on spatially compact manifolds is then completely described by the Gauss 
constraints which generate SU(2)-gauge transformations, the diffeomorphism 
constraints which generate spatial diffeomorphisms on $\Sigma$, and the 
Hamiltonian constraint, which is the generator of coordinate time 
evolution. In the non-compact case, true dynamics is generated by the boundary 
terms of the Hamiltonian. 
 
For compact spatial manifolds $\Sigma$, a well defined quantisation procedure 
for the above setup has been developed, which we review before discussing our 
extension to the non-compact case. The strategy is to specify an algebra of 
classical variables $\Aob$ and then to seek a representation of this algebra 
on some auxiliary Hilbert space $\Ha$. The second step is to obtain operator 
versions of the classical constraints and to then impose these on the Hilbert 
space to obtain a reduced space of physical states along with a representation 
of the subalgebra of observables that commute with the constraints. 
 
\subsection{The classical algebra of observables} 
 
To obtain the classical algebra of elementary functions which can be 
implemented in the quantum theory, we need to integrate the canonically 
conjugate variables, $\Aia$ and $e_{iab}$, against suitable smearing 
fields. In usual quantum field theory, these fields are 
three-dimensional. However, in canonical quantum general relativity, due to 
the absence of a background metric, it is more convenient to smear $n$-forms 
against $n$-dimensional surfaces instead of the usual three-dimensional 
ones~\cite{rovelli90,ashtekar92,ashtekar98}. 
 
Configuration observables can be constructed through holonomies of 
connections.  Given an embedded graph $\Gamma$ which is a collection of $n$ 
paths $\{\g[1],\ldots,\g[n]\} \in \Sigma$, and a smooth function $f$ from 
$\mathrm{SU(2)}^n$ to $\mathbb{C}$, we can construct cylindrical functions of 
the connection: 
\[ 
    \psi_{f,\Gamma}(A) = f(H(A,\g[1]),\ldots,H(A,\g[n])). 
\] 
$H(A,\g[i]) \in \mathrm{SU(2)}$ is the holonomy assigned to the edge $\g[i]$ 
of $\Gamma$ by the connection $A \in \A$. We denote by $\cyl$, the algebra 
generated by all the functions of this form. This is the space of 
configuration variables. To obtain momentum variables, we smear the two-forms 
$e_{abi}$ against distributional test fields $t^i$ which take values in the 
dual of $\mathfrak{su}(2)$ and have two dimensional support. This gives us 
\[ 
    E_{t,S} \equiv \int_{S} e_{abi} t^i dS^{ab}, 
\] 
where $S$ is a two-dimensional surface embedded in $\Sigma$. More 
precisely~(c.f.~\cite{ashtekar98}), we require that \mbox{$S = \bar{S} - 
\partial\bar{S}$}, where $\bar{S}$ is any \emph{compact}, analytic, two 
dimensional submanifold of $\Sigma$. 
 
The elements of $\cyl$ and the functions $E_{t,S}$ are the variables that we 
wish to promote to quantum operators. They form a large enough subset of all 
classical observables in the sense that they suffice to distinguish phase 
space points.  The algebra of elementary observables, $\Aob$, is the algebra 
generated by the cylindrical functions and the momentum variables, with the 
choice of Poisson brackets as given in~\cite{ashtekar98}. 
 
\subsection{The standard representation} \label{sec-rep} 
 
The next step in the quantisation procedure is to construct a Hilbert space 
$\Ha$ on which the algebra of elementary variables $\Aob$ is represented. In 
this subsection, we describe the construction of this Hilbert 
space~\cite{ashtekar95,baez95} concentrating on the GNS 
(Gel'fand-Naimark-Segal) construction since we shall later make crucial use of 
this technique. 
 
The GNS construction (see, e.g.,~\cite{haag93,landsman87,kadison83} for more 
detailed expositions) allows us to construct a representation of any 
$^*$-algebra $\Al$ for any given positive linear form (also called a state) 
$\om$ on this algebra. This is done in three steps: 
\begin{enumerate} 
    \item Using $\om$, define a scalar product on $\Al$, regarded as a
linear space over $\mathbb{C}$, by \[ \ip{\ba}{\bb} = \om(\ba^*\bb),
\] for $\ba,\bb \in \Al$. The positivity of $\om$ implies
$\ip{\ba}{\ba} \geq 0$.
 
    \item To obtain a positive definite scalar product, we construct
the quotient $\Al/\I$ of $\Al$ by the null space $\I = \{\ba \in \Al |
\om(\ba^*\ba) = 0\}$. We denote the equivalence classes in $\Al/\I$ by
$[\ba]$ and we have: \[ \ip{[\ba]}{[\ba]} \equiv \norm{\ba}^2 > 0 \] The
completion of $\Al/\I$ in the above norm is the carrier Hilbert space
$\Ho$ for our representation.
 
    \item Finally it can be shown that a representation $\pi_{\om}$ of
$\Al$ on $\Al/\I$ (which, if $\Al$ is a Banach $^*$-algebra, can be
extended continuously to $\Ho$) is given by: \[ \pi_{\om}(\ba)\Psi =
[\ba\bb], \] for $\Psi = \bb \in \Ho$ and $\ba \in \Al$.
\end{enumerate} 
 
Let us now return to our particular problem. We start by constructing a 
multiplicative representation of $\cyl$ on a Hilbert space $\Ha$, on which 
momentum operators will be shown to act as derivations. We use the GNS 
construction to construct the representation of $\cyl$.  To obtain a greater 
degree of control one first introduces a sup norm to complete $\cyl$ to a 
$C^*$-algebra $\ocyl$: 
\[ 
    \snorm{\psi_{f,\Gamma}} = \sup_{A \in \A} |\psi(A)_{f,\Gamma}|. 
\]

The key to representing $\cyl$ is to define a positive linear form on it. This 
can be done using the Haar measure $dg$ on SU(2) as follows\cite{ashtekar93} 
\begin{equation}\label{def-meas} 
    \om(\psi_{f,\Gamma}) = \int d\mu(A)\, \psi_{f,\Gamma} \equiv 
    \int_{\mathrm{SU(2)}^n} dg_1 \cdots dg_n\, f(g_1,\ldots,g_n), 
\end{equation} 
where $g_i \in \mathrm{SU(2)}$. Note that the right hand side does not depend 
on $\Gamma$. Nevertheless, our definition makes sense since, if\footnote{ This 
holds if $\Gamma$ consists of analytic paths but extensions to the 
non-analytic case are possible, c.f.~\cite{baez95a}.} $\psi_{f,\Gamma} = 
\psi'_{f',\Gamma'}$, then $\om(\psi) = \om(\psi')$. This allows us to define 
the (\underline{s}tandard) inner product: 
\begin{equation} \label{def-ip} 
    \ip{\psi_1}{\psi_2}_s = \om(\psi_1^*\psi_2) = 
    \int_{\mathrm{SU(2)}^n}f^*_1(g_1,\ldots,g_n)f_2(g_1,\ldots,g_n) dg_1\cdots 
    dg_n. 
\end{equation} 
Here we make use of the fact that if the functions $f_1$ and $f_2$ have a 
different number of arguments, say $f_1:\mathrm{SU(2)}^{m} \ra \mathbb{C}$ 
with $m < n$, we can trivially extend $f_1$ to a function on 
$\mathrm{SU(2)}^n$, which does not depend on the last $n-m$ arguments. Since 
this product is already positive definite, we can proceed directly with the 
completion of $\ocyl$ to obtain our auxiliary Hilbert space $\Ha$ carrying a 
multiplicative representation of the algebra $\cyl$. $\Ha$ can also be 
regarded as space of square integrable functions defined with respect to a 
genuine measure on some completion $\bar{\A}$ of $\A$ as is done 
in~\cite{ashtekar95}. 
 
We are left with the task of representing the momentum variables on this 
Hilbert space. This done by constructing essentially self adjoint operators 
$\hat{E}_{t,S}$ on $\cyl$ which can be extended to $\Ha$. These operators are 
derivations on $\cyl$ i.e.\ linear maps satisfying the Leibnitz rule, which 
act on functions $\psi_{f,\Gamma}\in \cyl$ only at points where $\Gamma$ 
intersects the oriented surface $S$. The precise definition of these operators 
is not needed for our purposes, but it can be found, e.g., 
in~\cite{ashtekar98}. This choice of operators gives the correct 
representation of the classical algebra $\Aob$, which provides us with a 
kinematical framework for canonical quantum gravity. In the following
this representation will be referred to as the standard representation
$\pi_s$. To obtain physical states 
we need to introduce the quantum constraints and study their action. 
 
\subsection{The constraints}\label{constraints} 
 
The simple geometrical interpretation of the Gauss and diffeomorphism 
constraints allows us to bypass the attempt to construct the corresponding 
constraint operators. Instead, we can construct unitary actions of the gauge 
group $\G$ and the diffeomorphism group $\D$ on $\Ha$ and demand that physical 
states be invariant under these actions. The imposition of the Hamiltonian 
constraint is still an open issue and we will not discuss it. In this sense, 
our entire discussion is at the kinematical level. 
 
The group $\G$ has a natural action on the space of connections which induces 
a unitary action of $\G$ on $\Ha$. Gauge invariance is simply achieved by 
restricting to the subspace $\Hg \subset \Ha$ of gauge invariant functions. It 
can be shown that this space is spanned by the so-called spin networks 
states~\cite{rovelli95a,baez95}. 
 
If we try to follow the same procedure for the diffeomorphism constraint, we 
find that there are no non-trivial diffeomorphism invariant states in $\Hg$. 
This problem is overcome by looking for distributional solutions to the 
constraints.  Again the natural pull-back action of the diffeomorphism group 
on the connections induces a unitary representation of $\D$ on $\Ha$ because 
of the diffeomorphism invariance of the inner product~(\ref{def-ip}).  One 
then considers a Gel'fand triple construction $F \subset \Hg \subset F'$, 
where $F$ is a dense subspace\footnote{ $F$ is usually chosen to be 
$\overline{\cyl}$.} of $\Hg$ and $F'$ its topological dual and identifies the 
reduced Hilbert space $\Hk$ with a subspace of $F'$ that is invariant under 
the dual action of the diffeomorphism constraint.  $\Hk$ carries a natural 
dual action of the algebra $\alk$, which is the subalgebra of $\Aob$ 
containing the elements that commute with the constraints. 
 
\subsection{Problems}\label{problems} 
 
As we stated in the introduction, we want to study states which
represent asymptotically flat classical metrics, especially Minkowski
space. We now argue that such states generically do not lie in the
Hilbert space $\Ha$ constructed above, which is the reason that this
representation is not adequate for non-compact $\Sigma$.
 
States which describe non-compact geometries should either be based on
curves of infinite length or an infinite number of curves. This is
because the area and volumes of regions of $\Sigma$ which do not
contain edges and vertices of graphs vanish.  A particular example of
an attempt to construct a state which approximates a chosen flat
Euclidean 3-metric $g_{ab}$ on $\Sigma$ is the so-called weave given
in~\cite{grot97}.  This weave is based on an infinite collection of
graphs $\Gamma_{r,\mu} = \bigcup_{i=1}^{\infty} D_i$, where $D_i$ is
the union of two randomly oriented circles $\gamma_i^a$ and
$\gamma_i^b$ of radius $r$, which intersect in one point. To ensure
isotropy these graphs are sprinkled randomly in $\Sigma$, with
sprinkling density $\mu$, where $\mu$ is defined with respect to
$g_{ab}$.  Given $n$ double circles $D_i$ we can construct the
cylindrical function $\W_n$:
\begin{equation}\label{weave} 
    \W_n(A) = \prod_{i=1}^n 
    \mathrm{Tr}[\rho_1(H(\gamma_i^a,A)H(\gamma_i^b,A))] 
\end{equation} 
where $\rho_1$ denotes the fundamental representation of SU(2).  The weave 
state $\W$ should arise in the limit $n \rightarrow \infty$. This limit does 
not exist in the Hilbert space $\Ha$, since the above sequence $\W_n$ is not 
Cauchy in either the sup norm or the $L^2$ norm based on the inner product 
(\ref{def-ip}). This holds even if we impose physically reasonable 
(non-uniform) fall-off conditions on the connections, such as those for 
asymptotically flat gravity. The basic problem is that for any curve embedded 
in $\Sigma$ we can always find a connection that will assign to this curve any 
holonomy we choose~\cite{ashtekar93}\footnote{We thank John Baez for this observation.}. In 
particular, this means that if we have a state $f(H(\gamma,A))$ then $\sup 
|f(H(\gamma,A))|$ will be independent of the location of $\gamma$. This is a 
generic result, and we conclude that a large class of physically interesting 
states based on infinite collections of graphs do not exist in $\Ha$. Similar 
arguments can be used to show that states based on curves of infinite length 
do not lie in $\Ha$ either. 
 
A very natural way of dealing with states based on a finite number of curves of 
infinite length was introduced in~\cite{arnsdorf99}. The key point is to 
consider a compactification of $\Sigma$ and show that these states 
belong to the 
auxiliary Hilbert space constructed on the compactified manifold. Problems 
arise when trying to extend this approach to discuss cylindrical functions 
based on graphs with infinite number of edges and vertices as cluster points of 
vertices necessarily arise in the compactified manifold. This just illustrates 
the fact that the Hilbert space $\Ha$, along with the representation of 
observables it carries, was constructed for compact spatial slices and is not 
adequate to describe the case of non-compact $\Sigma$. In the next section, we 
propose a different solution to the above problems by giving a procedure to 
construct new Hilbert spaces for quantum general relativity, which 
describe fluctuations around specified background states. In 
particular, these states can have non-compact support on the spatial manifold.

\subsection{A new representation for $\Aob$} \label{sec-newrep} 
 
We take an approach analogous to that in algebraic field theory, where 
Hilbert spaces describing field theory at finite temperature arise as 
inequivalent representations of the algebra of observables via the GNS 
construction. In this approach, the algebra of observables is 
considered as primary, as opposed to the Hilbert space of states. This 
gives us the flexibility to consider different Hilbert spaces 
depending on which background state we are interested in. The vectors 
in this space then describe finite perturbations around this preferred 
state. In practice, we use the background state to define a positive 
linear form on our observable algebra $\Aob$ by interpreting the form as 
the expectation value of the observables in the preferred state. This 
then gives the starting point for the GNS construction which leads to 
the desired quantum theory. We have the following procedure: 

\begin{enumerate} 
\item The fact that we have a representation $\pi_s$ of $\Aob$ on
$\Ha$ as given in section~\ref{sec-rep} enables us to identify $\Aob$
 with a subalgebra of the concrete $^*$-algebra\footnote{If one
 desires to work with a $C^*$-algebra one faces the problem that the
 `momentum' operators are unbounded. To proceed one needs to consider
 algebras of bounded functions on $\Aob$ or consider families of
 spectral projectors of the unbounded operators. Physically, this does
 not lead to a loss of generality, c.f.~\cite{haag93}.}  of operators on $\Ha$
 
\item Now we define a new positive linear form $\om$ on $\Aob$, which 
is interpreted as the `vacuum' expectation value of the elementary 
variables. Note that that in contrast to section~\ref{sec-rep} we are 
defining $\om$ on all of $\Aob$ not just $\cyl$. 
 
\item Using the GNS construction we can now proceed to construct a 
representation of $\Aob$. The vectors in the carrier Hilbert space will 
be equivalence classes of elements of $\Aob$, which should be 
interpreted as excitations of the `vacuum' state, obtained by 
acting on the `vacuum' 
with the corresponding algebra element. 
\end{enumerate} 
 
In the remainder of this paper we will demonstrate how we can 
construct $\om$ explicitly. We do this by constructing a modified 
weave, which approximates flat space. 
As before the weave is not well defined as a state in $\Ha$ but 
it can be used to define $\om$, which will give the expectation values
of the elements of $\Aob$ in this background state.

\section{Approximating Classical Geometries}\label{sec-weave}

Let us start with a brief discussion of the problem of approximating classical 
metrics by quantum states. It is generally accepted that to obtain 
semi-classical behaviour from a quantum theory, one needs two things: i) a 
suitable coarse-graining, and, ii) coherent states. So far, however, coherent 
states have not been constructed in loop quantum gravity. To obtain these one 
needs to construct a state in which neither the 3-geometry nor its 
time-derivative are sharp. Rather, they should both have some minimum spreads 
as dictated by the uncertainty principle. We will come back to this point 
later. 
 
An example of states which approximate classical 3-metrics are weave 
states (well-defined only if $\Sigma$ is compact), such as the one 
defined in eq.~(\ref{weave}) (see 
e.g.~\cite{grot97,ashtekar92}). However, all weaves which have 
been constructed so far are eigenstates of the 3-geometry, so they are 
highly delocalized in their time derivative. Intuitively, this means 
that while a weave may approximate the 3-metric at one instant of 
time, evolving the state for even an infinitesimal time will 
completely destroy this approximation. In this section, we will 
construct a more satisfactory set of states that can also be used to 
approximate 3-metrics. In particular, these states can be used to 
define a positive linear form on $\Aob$ as is needed for the GNS 
construction even in the case that we want to approximate a 
non-compact geometry. 
 
\subsection{Approximating 3-metrics} 
 
Let us now take a closer look at the weaves. Their construction is
made possible by the existence of operators on $\Ha$ which measure the
area of a surface and the volume of a
region~\cite{rovelli95,depietri96,ashtekar97,ashtekar97a}. This allows
us to approximate classical metrics by requiring that the expectation
values of areas and volumes of macroscopic surfaces and regions agree
with the classical values.
 
For concreteness, in the rest of this section, we shall restrict
ourselves to the problem of approximating the flat Euclidean metric on
$\mathbb{R}^3$. Let $w$ be a state which approximates the flat space
at scales larger than a cut-off scale $l_c$. The approximation problem
can then be stated as follows: Given any object (of characteristic
size larger than $l_c$) with bulk $R$ and surface $S$ in
$\mathbb{R}^3$, we wish to make repeated measurements of the volume of
the region $V[R]$ and the area of the surface $A[S]$ in the state $w$
while placing the object at different points in space. If we wish to
recover values corresponding to the flat metric $g_{ab}$ on
$\mathbb{R}^3$ at large scales, we require:
\begin{enumerate} 
    \item The average values of the area and volume for $S$ and $R$ obtained 
    during the measurements should be given by the classical values: 
    \begin{eqnarray} 
        \overline{\langle A_S\rangle} &=& A_g[S] \equiv \int_S \sqrt{\det 
        {}^2g_{ab}}\label{eigenarea}\\ 
        \overline{\langle V_R\rangle} &=& V_g[R] \equiv \int_R \sqrt{\det 
        g_{ab}}\label{eigenvol} 
    \end{eqnarray} 
    where ${}^2g_{ab}$ denotes the induced 2-metric on $S$. Here a bar over the 
    value indicates an average with respect to position in space whereas the 
    angle brackets indicate the expectation value in the quantum state. 
 
    \item The standard deviation $\sigma$ of the measurements should be 
    small compared to the length scale $\ell_c$ accessible by current 
    measurements: 
    \begin{equation}\label{deviation1} 
        \sigma_{V} \ll \ell_c^3 \qquad {\rm and} \qquad 
        \sigma_{A} \ll \ell_c^2, 
    \end{equation} 
    where $\sigma_{V}$ and $\sigma_{A}$ denote the standard deviations 
    in a series of measurements determining the volume and area of 
    objects of the scale $\ell_c$. 
\end{enumerate} 
 
We shall call any state satisfying conditions (\ref{eigenarea}), 
(\ref{eigenvol}) and (\ref{deviation1}), a weave state. These 
conditions do not determine a state uniquely. Rather, one can 
construct infinitely many states which satisfy them. Below, we give an 
example of a weave state which is not an eigenstate of the 
three-geometry, but is peaked in both the connection and spin-networks 
bases. 
We refer to this state as a ``quasi-coherent'' 
weave in the following. 
 
\subsection{A ``quasi-coherent'' weave, $\Q$} 
 
Since a weave has to give areas and volumes to all surfaces in a 
non-compact manifold, the state must be based on an infinite graph. 
We take this graph to be $\Gamma_{r,\mu}$ as defined in section~\ref{problems}. 
The values of the parameters $r$ 
and $\mu$ will be determined by the requirement that the state based on 
$\Gamma_{r,\mu}$ satisfy the weave conditions. 
 
To define the state, we start with the cylindrical function $q_i$ 
based on the graph $D_i$ in $\Gamma_{r,\mu}$: 
\begin{equation}\label{weave-fn} 
    q_i(A) = \eta \exp \left(\lambda\mathrm{Tr}\left[\rho_1(H(\gamma_i^a,A) 
    H(\gamma_i^b,A)) - \rho_1(e)\right]\right), 
\end{equation} 
where $\lambda$ is an arbitrary constant, $e$ is the identity in SU(2) and 
$\eta$ is a normalisation factor. We also consider the products: 
\[ 
    \Q_n(A) = \prod_{i=1}^n q_i(A). 
\] 
The state that we are interested in, is the limit $\Q = \Q_{\infty}$, which is 
again not an element of the standard Hilbert space. Nevertheless, as 
we will 
see, this product can serve as a background  
for the construction of new 
Hilbert spaces, describing excitations of $\Q$. 
 
Let us denote the connection that gives a trivial holonomy on all paths by 
$A_0$. By construction the functions $\Q_n(A)$ take on their maximum values at 
$A_0$. Conversely, knowledge of the holonomies on all paths in $\Sigma$ allows 
us to determine a corresponding connection uniquely. Hence as $n \ra \infty$ 
the function $\Q_n$ becomes increasingly peaked around $A_0$, the sharpness of 
the peak being determined by $\lambda$. This is one of the reasons for calling 
our weave a ``quasi-coherent'' state, the other being the exponential 
dependence on group elements which is characteristic of coherent states. We 
will explore these properties of the state further in future work. For the 
present, we are interested in showing that $\Q$ is a good weave. In order to do 
so, we need to show that it satisfies the weave conditions (\ref{eigenarea}), 
(\ref{eigenvol}) and (\ref{deviation1}). We shall do this by demonstrating that 
standard deviations of area and volume measurements are roughly of the order of 
$\ell_c\ell_P$ and $\sqrt{(\ell_c\ell_P)^3}$, where $\ell_P = \sqrt{\hbar 
G_{\rm Newton}/c^3}$ is the Planck length and hence much smaller than the 
bounds set by eq.\ (\ref{deviation1}). If only interested in how $\Q$ 
can be used to construct new Hilbert spaces the reader may skip to the 
next section.  
 
\begin{figure}[t] 
    \begin{center} 
        \psfrag{l1}{\small $\lambda=10$} 
        \psfrag{l2}{\small $\lambda=30$} 
        \psfrag{l3}{\small $\lambda=50$} 
        \psfrag{sp2}{\small $s_p^2$} 
        \psfrag{p}{\small $p$} 
        \psfrag{0}{\small $0$} 
        \psfrag{5}{\small $5$} 
        \psfrag{10}{\small $10$} 
        \psfrag{15}{\small $15$} 
        \psfrag{20}{\small $20$} 
        \psfrag{25}{\small $25$} 
        \psfrag{30}{\small $30$} 
        \psfrag{0.00}{\small $0.00$} 
        \psfrag{0.04}{\small $0.04$} 
        \psfrag{0.08}{\small $0.08$} 
        \psfrag{0.12}{\small $0.12$} 
        \psfrag{0.16}{\small $0.16$} 
        \psfrag{0.20}{\small $0.20$} 
        \includegraphics[height=2.5in,keepaspectratio,clip]{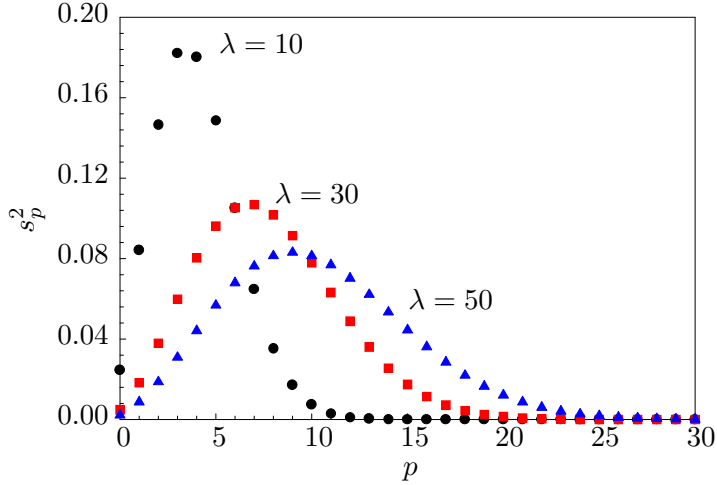} 
    \end{center} 
    \caption{Normalised coefficients $s_p^2$ for $\lambda=10$, $\lambda = 30$ 
    and $\lambda=50$. The sharpness of the peak decreases with increasing 
    $\lambda$.}\label{coeffs} 
\end{figure}

Let us start by expanding the state $q_i(A)$ in into an eigenbasis of the area 
operator and calculate the area expectation values and deviations. We do this 
by noting that the cylindrical function $f_p(A) = 
\mathrm{Tr}[\rho_p(H(\gamma_i^a,A)H(\gamma_i^b,A))]$ based on the graph $D_i$ 
is an eigenstate of the area operator\footnote{This follows since $f_p$ can be 
expanded in terms of spin-network functions that all assign $\rho_p$ to each 
edge of $D_i$. Hence all spin-network functions in the expansion have the same 
area eigenvalues.},where $\rho_p$ is a representation of SU(2) in `colour' 
notation, i.e.\ $p=\dim(\rho) -1$.  The eigenvalues $a_p$ of the area operator 
corresponding to some surface $S$, which intersects $D_i$ \emph{exactly once}, 
are given by $16\pi \ell_P^2\sqrt{\frac{p}{2}(\frac{p}{2} +1)}$. Thus, to 
evaluate the area expectation value $\area$ and the deviation $\adev$ of this 
operator we start by expanding the function $q_i(A)$ in terms of the area 
eigenstates $f_p(A)$: $q_i = \sum_p s_p f_p$. We want to determine the 
coefficients of this expansion. We start by noting that $q_i$ is defined by its 
series expansion (the $i$ index labelling the graph will be suppressed in the 
following): 
\[ 
    q = \eta e^{-2\lambda}(1 + \lambda f_1 + \frac{\lambda^2f_1^2}{2!} + 
    \frac{\lambda^3f_1^3}{3!} + \cdots). 
\] 
Hence, we need to expand $f_1^n$ in terms of $f_p$ to determine the $s_p$'s. 
This can be done by using the decomposition rules for tensor products of 
representations of SU(2): 
\[ 
    \mathrm{Tr}^n[\rho_1(g)] = \mathrm{Tr}[\oplus_p c_p^n \rho_p(g)]. 
\] 
Hence, it follows that $f^n_1 = \sum_p c^n_p f_p$. To determine the 
coefficients $c^n_p$, we use the fact that $\rho_p \otimes \rho_1 = \rho_{p-1} 
\oplus \rho_{p+1}$ and that: 
\[ 
    \underbrace{\rho_1 \otimes \ldots \otimes \rho_1}_n = (\oplus c^{n-1}_p 
    \rho_p)\otimes \rho_1. 
\] 
This gives us the following recursion relation: 
\[ 
    c^n_p = c^{n-1}_{p-1} + c^{n-1}_{p+1}, \qquad {n,p \ge 0, \ p \leq n}. 
\] 
Using the condition $c_0^0=1$, we can solve this recursion relation to get: 
\begin{equation}\label{cnp} 
    c^n_p  =  \frac{(p+1)n!}{(\frac{n-p}{2})!(\frac{n+p}{2}+1)!} \qquad 
   \mathrm{for}\ \frac{n-p}{2} \in \mathbb{N},\\ 
\end{equation} 
and $c^n_p = 0$ otherwise. The expansion coefficients $s_p$ (as a function of 
$\lambda$) are given by 
\[ 
    s_p(\lambda) = \mathcal{N}(\lambda) \sum_{n = 0}^{\infty}\frac{\lambda^n 
    c^n_p}{n!} 
\] 
where $\mathcal{N}(\lambda)$ is defined such that $\sum_{p=0}^{\infty} s_p^2 = 
1$. Substituting eq.~(\ref{cnp}) into the right hand side of the above 
expression, we find: 
\begin{eqnarray*} 
    s_p(\lambda) & = &\mathcal{N}(\lambda) (p+1)\sum_{k=0}^\infty 
    \frac{\lambda^{2k+p}}{k!(k+p+1)!} =  \mathcal{N}(\lambda) \frac{(p+1)}{\lambda} 
    I_{p+1}(2\lambda)\\ 
    \mbox{} & = & \mathcal{N}(\lambda) [I_p(2\lambda) - I_{p+2}(2\lambda)], 
\end{eqnarray*} 
where $I_p(x)$ is the modified Bessel function of order $p$. Using the 
properties of Bessel functions, we can then evaluate the normalisation constant 
to be 
\[ 
   \mathcal{N}(\lambda) = [I_0(4\lambda) - I_2(4\lambda)]^{-1/2}. 
\] 
Figure~\ref{coeffs} shows the numerical values of the coefficients 
$s_p^2$ for a few values of $\lambda$. 
 
The final form of the expansion of $q_i$ in terms of the area eigenstates $f_p$ 
is 
\begin{equation} 
    q(A) = \sum_{p=0}^\infty \frac{I_p(2\lambda) - I_{p+2}(2\lambda)}{ 
    \sqrt{I_0(4\lambda) - I_2(4\lambda)}}\, f_p(A). 
\end{equation} 
We needed to show that the states $q_i$ are peaked in area and volume. Let us 
consider, in particular, $\lambda=30$. In this case, we have: 
\[ 
    \area = 4.33 (16\pi \ell_P^2) \qquad \mathrm{and} \qquad \adev \equiv 
    \sqrt{\langle a^2 \rangle - \area^2} = 1.87 (16\pi \ell_P^2), 
\] 
and hence both values are of order $\ell_P^2$. The explicit calculation for the 
volume expectation value and deviation, $\vol$ and $\vdev$, in the state $q_i$ 
is somewhat more complicated since the states $f_p$ have to be expanded in 
volume eigenstates. After doing this, we find: 
\[ 
    \vol = 3.61 (16\pi\ell_P^2)^{3/2}  \qquad \mathrm{and} \qquad \vdev \equiv 
    \sqrt{\langle v^2 \rangle - \vol^2} =  2.21 (16\pi \ell_P^2)^{3/2}. 
\] 
Again, both values are of the order of a few $\ell_P^3$. 
 
In general, the area expectation value for the area as a function of $\lambda$ 
can be written as 
\[ 
    \area = 16\pi \ell_P^2 \sum_{p=0}^\infty \frac{(I_p(2\lambda) - 
    I_{p+2}(2\lambda))^2} {I_0(4\lambda) - I_2(4\lambda)} 
    \sqrt{\frac{p}{2}\left(\frac{p}{2} + 1 \right)}. 
\] 
We do not have a closed form for $\vol$ since the evaluation of the volume 
expectation value involves the diagonalization of a matrix. However, since the 
largest eigenvalue of the volume operator in the state $f_p$ increases with $p$ 
as $p^{3/2}$, for a value of $\lambda$ for which the expansion is dominated by 
a few small $p$'s, we can see that $\vol$ has to be of the order of a few 
$\ell_P^3$. Given the fact that  both $\area$ and $\vol$ have value of the 
order of a few Planck units, it is reasonable to assume that the 
bound~(\ref{deviation1}) is satisfied for area and volumes in the state $\Q$, 
which is a product of the $q_i$'s, as well. We now show that this is indeed the 
case. 
 
The expectation value of the volume operator for a region $\hat{V}_R$ depends 
upon the number $N_R$ of double circles  in the region $R$, namely: 
\[ 
    \langle V_R \rangle = N_R \vol. 
\] 
Similarly, the expectation value for the area operator of a surface $\hat{A}_S$ 
depends on the number of intersections $N_S$ of the double circles with the 
surface: 
\[ 
    \langle A_S \rangle = N_S \area. 
\] 
Because the double circles $D_i$ are sprinkled randomly in $\Sigma$ the number 
of the graphs in any given region $R$ of volume $V_g[R]$ is given by a Poisson 
distribution. In particular, the average value is given by $\bar{N}_R = 
V_g[R]\mu$. Hence average values of the above measurements are given by: 
\begin{equation}\label{avarea} 
    \overline{\langle V_R \rangle} = \bar{N}_R \vol 
\end{equation} 
and 
\begin{equation}\label{avvol} 
    \overline{\langle A_S \rangle} = \bar{N}_S \area, 
\end{equation} 
where $\bar{N}_S = \frac{2}{3}V[S]\mu$ and $V[S] = 6rA_g[S]$ denotes the volume 
of a shell surrounding $S$ with thickness $3r$ on either side. The factor $2/3$ 
is the average number of crossings between a double circle within this shell 
with $S$ as determined in~\cite{grot97}. 
 
To determine the standard deviations $\Vdev$ and $\Adev$ around $\oVol$ and 
$\oArea$ we use an approximation: we assume that the eigenvalues of $\hat{V}_R$ 
and $\hat{A}_S$ are given by the product of two independent quantities, i.e.\ 
$N_R(V_R/\bar{N}_R)$ and $N_S(A_S/\bar{N}_S)$ respectively, where $V_R$ and 
$A_S$ are sums of any $n$ elementary eigenvalues $v_p$ and $a_p$. The 
deviations of the quantities $N_R$ and $N_S$ are given by $\sqrt{\bar{N}_R}$ 
and $\sqrt{\frac{2}{3}\bar{N}_S}$, since we are dealing with a Poisson 
distribution of graphs. Deviations of $V_R/\bar{N}_R$ and $A_S/\bar{N}_S$ on 
the other hand are $\vdev/\sqrt{\bar{N}_R}$ and $\adev/\sqrt{\bar{N}_S}$. Our 
approximation implies that: 
\begin{eqnarray*} 
    \Adev & \approx &  \sqrt{\bar{N}_S} 
    \sqrt{\adev^2 + \frac{2}{3}\area^2}\\ 
    \Vdev & \approx & 
    \sqrt{\bar{N}_R}\sqrt{\vdev^2 + \vol^2} 
\end{eqnarray*} 
 
From eqs.~(\ref{eigenarea}), (\ref{eigenvol}), (\ref{avarea}), and 
(\ref{avvol}), we find: $\bar{N}_S = A_g[S]/\area$ and $\bar{N}_R = 
V_g[R]/\vol$. Because of the scale of $\area$, $\adev$ and $\vol$, $\vdev$ we 
conclude that eq.~(\ref{deviation1}) is satisfied. Thus, we have shown that 
$\Q$ is a weave state. For any particular value of $\lambda$, we can determine 
$\mu$ and $r$ using eqs.~(\ref{avarea}), (\ref{avvol}), (\ref{eigenarea}) and 
(\ref{eigenvol}): 
\[ 
    \mu = \vol^{-1} \qquad \mathrm{and} \qquad r = (4\mu\area)^{-1}. 
\] 
We will denote the infinite collection $\bigcup_{i=1}^{\infty}D_i$ of double 
circles $D_i$ with $r$ and $\mu$ determined by the above conditions by $\gq$.

\section{New Hilbert Spaces} 
 
We now show how we can use $\Q$ to define a new representation of the 
algebra $\Aob$ following the steps outlined in section~\ref{sec-newrep}. 
We begin by noting that we have a representation $\pi_s$ of  
$\Aob$ on $\Ha$, but as we have seen $\Q$ 
does not belong to this Hilbert space. Nevertheless, we can define the 
action of an element of $\Aob$ on $\Q$. The crucial point is that the 
elementary quantum observables --- the elements of $\cyl$ and the 
derivations on $\cyl$ --- have support on a compact spatial region, 
which is a direct consequence of the smearing needed to make sense of 
the classical expressions. But if we restrict $\Q$ to any compact 
region of $\Sigma$ we obtain an element of $\Ha$ by restricting the 
underlying graph $\gq$ to that region. Hence given an arbitrary 
element $\ba \in \Aob$ we proceed as follows: 
 
\begin{enumerate} 
    \item Denote the closure of the support of $\ba$ by $R \subset \Sigma$. 

    \item Construct the graph $\gq|_R$ of $\gq$ restricted to $R$: 
    \[ 
         \gq|_R \equiv \bigcup_{D_i \cap R \neq \emptyset} D_i. 
    \]  
In other words, consider the union of all double circles which have a non-zero 
intersection with the support of $\ba$. This graph is finite, since $R$ 
is compact and the double circles $D_i$ are sprinkled in $\Sigma$ with finite 
density $\mu$, we obtain the state $\QR \in \Ha$ which is given by restricting 
$\Q$ to the graph $\gq|_R$: 
\[ 
     \QR \equiv \mathbb{I} \cdot \prod_{D_i \in \gq|_R} q_i, 
\] 
where $\mathbb{I}(A) = 1$ for all $A$ is the identity function.
This state has unit norm in $\Ha$ since all the $q_i$'s are normalised. 
 
    \item Since $\QR \in \Ha$, the action of $\ba$ on $\QR$ 
    denoted by $\pi(\ba)_s\QR$ is well-defined. It is understood 
    that the region $R$ will depend on $\ba$. 
\end{enumerate} 
 
This allows us to define $\om_{\Q}(\ba)$: 
\begin{equation}\label{form} 
    \om_{\Q}(\ba) = \ip{\Q|_R}{\pi_s(\ba)\Q|_R}_s = \int \QR^*\ \pi_s(\ba)
\QR \ d\mu(A),
\end{equation} 
where the integral is defined as in eq.~(\ref{def-meas}). This is
well-defined since the integrand is an element of $\Ha$. It follows
from the fact that equations~(\ref{def-ip}) defines a true inner
product $\ip{\cdot}{\cdot}_s$ on $\Ha$ that $\om$ is
indeed a positive (not necessarily strictly positive) linear form
on $\Aob$.

 
Given this positive linear functional, we could proceed with steps 2 
and 3 of the GNS construction outlined in section~\ref{sec-rep} to 
obtain a representation of the algebra $\Aob$. Instead, to get a more 
intuitive representation we make use of the theorem below to construct 
a unitarily equivalent representation $\pi_{\Q}$ of $\Aob$ on a 
Hilbert space $\Hq$ obtained by defining a new inner product on 
$\cyl$. 
\begin{theorem} 
    Any representation $\pi$ of a $^*$-algebra $\Al$ with cyclic
    vector\footnote{The vector $\chi \in \mathcal{H}$ is cyclic with
    respect to the representation $\pi$ of $\Al$ on $\mathcal{H}$ if
    $\pi(\Al)\chi$ is dense in $\mathcal{H}$.} $\chi$ such that: \[
    \ip{\chi}{\pi(\ba)\chi} = \om(\ba), \] for all $\ba \in \Al$ is
    unitarily equivalent to the GNS representation $\po$, with cyclic
vector $\chi_{\om}$ (corresponding to the unit element in $\Al$).
\end{theorem} 
 
\textbf{Proof} The proof of this theorem is analogous to the one of 
proposition 4.5.3 in~\cite{kadison83}. To proceed, we note that for each $\ba 
\in \Al$, 
\begin{eqnarray*} 
  \norm{\pi(\ba)\chi}^2 &=& \ip{\pi(\ba)\chi}{\pi(\ba)\chi} =
  \ip{\pi(\ba)\chi}{\pi(\ba^*\ba)\chi}\\ \mbox{}& = & \om(\ba^*\ba) =
  \norm{\ba}_\om^2
\end{eqnarray*} 
where $\norm{\cdot}_\om^2$ is the norm in the Hilbert space $\mathcal{H}_\om$ 
which carries the GNS representation. This means that there exists a 
norm-preserving, linear operator $U_0$ such that $U_0\pi(\ba)\chi = 
\po\chi_{\om}$. Since $\chi$ is a cyclic vector, U extends by continuity to an 
isomorphism from the Hilbert space $\mathcal{H}$ to $\mathcal{H}_\om$ and 
$U\chi_{\om} = \chi$. 
 
For any $\ba,\bb \in \Al$,  
\begin{eqnarray*} 
U \po(\ba)\po(\bb)\chi_{\om} & = & U \po(\ba\bb)\chi_{\om} =
\pi(\ba\bb) \chi\\
\mbox{} & = & \pi(\ba) \pi(\bb) \chi = \pi(\ba) U \po(b)\chi_{\om}. 
\end{eqnarray*} 
Since $\chi_{\om}$ is cyclic in $\mathcal{H}_\om$, we have $U\po(\ba) = 
\pi(\ba)U$ which in turn implies that $\pi(\ba) = U\po(\ba)U^*$. \qed 
 
Using the above theorem, we proceed as follows: 
\begin{enumerate} 
    \item On $\cyl$, the space of cylindrical functions, introduce the 
      following strictly positive inner product: 
    \begin{equation}\label{def-ip2} 
        \ip{\psi_{f_1,\Gamma_1}}{\psi_{f_2,\Gamma_2}}_{\Q} = \int \QR^*\QR 
        \psi_1^*\psi_2\ d\mu(A), 
    \end{equation} 
    where $R$ here is the union of the graphs 
    $\Gamma_1$ and $\Gamma_2$. Completion of $\cyl$ with respect to this 
    positive definite inner product gives us the Hilbert space $\Hq$. 
 
    \item We construct a representation $\pi_{\Q}$ of $\Aob$ on 
    $\cyl$, which is dense in $\Hq$ by:
 \begin{equation}\label{def-rep2} 
    \pi_{\Q}(\ba)\psi = \QR^{-1} \pi_s(\ba) (\QR\psi),
 \end{equation}
where 
    $\ba\in\Aob$, $\psi\in\cyl$ and $\QR^{-1}$ denotes the inverse 
    function of $\QR$ i.e., $\QR^{-1} \QR = \mathbb{I}$.  At this 
    point we note an additional requirement for the background state: 
    \mbox{$\Q(A) \neq 0$} for all $A$.  This invertability property is 
    motivated physically since our background state is meant to 
    represent an infinite `condensate of gravitons'. We should be able 
    to annihilate as well as create these gravitons, which motivates 
    invertability. The definition (\ref{weave-fn}) of $q_i(A)$ was chosen  
    to satisfy this property. 
\end{enumerate} 
 
It is now straightforward to see that $\pi_{\Q}$ is unitarily 
equivalent to the GNS representation defined via $\om_{\Q}$ given in
equation~(\ref{form}). We note that $\mathbb{I}$ is a 
cyclic vector in $\Hq$ with respect to the representation $\pi_{\Q}$, 
since $\Hq$ is the closure of $\cyl$.  In addition we have: 
\[ 
    \ip{\mathbb{I}}{\pi_{\Q}(\ba)\mathbb{I}}_{\Q} = \int \QR^*\QR 
\QR^{-1} \pi(\ba)_s  
    \QR \ d\mu(A) = \om_{\Q}(\ba), 
\] 
for all $\ba\in\Aob$. 
 
Hence we have constructed a Hilbert space and representation of
observables on it that describes fluctuations restricted to
essentially compact regions around some fixed infinite background
state. Intuitively, the above representation has a clear
interpretation. It acts on states by composing them with the
background.  We can regard the algebra of cylindrical functions $\cyl$
as creating and annihilating excitations on the background state. This
new representation $\pi_{\Q}$ on $\Hq$ is related to the standard
representation $\pi_s$ on $\Ha$ given in section~\ref{sec-rep} by
equation~(\ref{def-rep2}), i.e.: $\Q|_R\pi_{\Q}(\ba) =
\pi_s(\ba)\Q|_R$. Since $\Q|_R$ depends on the algebra element $\ba$,
which is a direct consequence of $\Q$ not being an element of $\Ha$,
this does not give us a unitary map between Hilbert spaces, instead we
have many different maps. It follows that
the
representations $\pi_{\Q}$ and $\pi_s$ are unitarily
\emph{inequivalent}\footnote{
This can be also be seen by considering
the following example. The state $\mathbb{I} \in \Ha$ has the property
that it is annihilated by \emph{all} ``momentum'' operators. If there
were a unitary map $U:\Ha \ra \Hq$ then $U\mathbb{I} \equiv \Psi
\in \Hq$ should have the same property, i.e.:
$\Q|_{\bar{S}}^{-1}\pi_s(E_{t,S})(\Q|_{\bar{S}} \Psi) = 0$ for all
derivations $E_{t,S}$. But since $\Psi$ has essentially compact
support in $\Sigma$ and since we can chose $S$ so that
$\pi_s(E_{t,S})\Q|_{\bar{S}} \neq 0$ outside any compact region,
this cannot be satisfied for all regions $S$. 
}.
The construction we have presented is very general and can be applied
to a large class of background states provided they satisfy the
necessary invertability condition.

\subsection{Constraints and Asymptotic Symmetries} 
 
We can proceed to reduce the Hilbert space obtained in the previous
section by imposing the Gauss and diffeomorphism constraints. In
general, when considering GNS representations $\pi$ of an algebra
$\Al$ we can implement actions of symmetry groups $G$ using the
following theorem (eq.\ III.3.14 in~\cite{haag93}):
\begin{theorem}\label{SymmThm} 
Given an action of $G$ on $\Al$: $\ba \ra g\ba$ such that $\om(g\ba)=
\om(\ba)$ for all $\ba \in \Al$ and $g \in G$ then we can define a
unitary representation $U$ of $G$ on $\Ho$ by:
\[ 
    U(g)\pi(\ba)\chi_{\om} = \pi(g\ba)\chi_{\om} 
\] 
where $\chi_{\om}$ is a cyclic vector in $\Ho$. 
\end{theorem}

In practice, when considering the representation $\pi_{\Q}$ we can
proceed as in section~\ref{constraints} to reduce the Hilbert space
$\Hq$. As before, there is no problem in implementing the Gauss
constraint. Since the state $\QR$ is invariant under gauge
transformations, the inner product defined in eq.~(\ref{def-ip2})
is also gauge invariant and we have a unitary action of the gauge
group on the state space. Again we implement the Gauss constraint by
restricting to the subspace of $\Hq$ consisting of gauge invariant
states, which is spanned by spin-networks.
 
When considering diffeomorphisms, we notice that $\QR$ is invariant only under 
diffeomorphisms that leave the graph $\gq$, on which $\Q$ is based, invariant. 
Let us denote this subgroup of diffeomorphisms by $\Dg$. Invariance of the 
inner product under $\Dg$ gives us a unitary representation of this group and 
we can use the Gel'fand triple construction detailed earlier to obtain a space 
of kinematical states $\Hqk$ that is invariant under $\Dg$. This space then 
naturally carries a dual representation of the subalgebra $\alk \subset \Aob$ 
of operators that commute with constraints. 
 
To discuss the significance of the breaking of diffeomorphism 
invariance to the group $\Dg$, we note that given two different 
background states $\Q$ and $\Q'$ which are defined with respect to 
diffeomorphic graphs: $\Gamma_{\Q'} = \phi 
\circ \gq$, where $\phi$ is any diffeomorphism, we obtain unitarily 
equivalent representations of 
$\Aob$: 
\[ 
    U[\pi_{\Q}(\ba)\mathbb{I}] = \pi_{\Q'}(\phi \ba)\mathbb{I}, 
\] 
where unitarity of $U$ follows from the diffeomorphism invariance of the inner 
product given by~(\ref{def-ip}). Hence, the choice of a particular member of 
the diffeomorphism class of $\Q$ is simply a partial gauge fixing. A 
physically equivalent way of getting the same results would be to average over 
the group $\Dg$. 
 
To conclude, we note that in the context of asymptotically flat
general relativity, which is the prime case of interest involving
non-compact spatial manifolds, the invariance group of the theory is
restricted to the connected component of the asymptotically trivial
diffeomorphisms. In the neighbourhood of infinity we would like to
have a unitary action of the Poincar\'{e} group on our state
space. Since physically relevant operators are typically evaluated at
infinity, this invariance is what is of prime interest. 
 
We shall discuss the construction of the action of the full Poincar\'e
group in future work. Here, we show how a unitary action of the
Euclidean group $E$ acting on $\Sigma$ can be incorporated in
our scheme. 
From theorem
\ref{SymmThm}, it follows that to do this we need a linear form on the algebra 
of observables that is invariant under the action of $E$. Such a form
can be obtained by using the fact that the Euclidean group is locally
compact to group average the form $\om$ given in
eq.~(\ref{form}). Given an increasing sequence of compact subsets $S_k
\subset E$, $S_k \subset S_{k+1}$, $\cup S_k = E$ we define:
\[ 
    \om_k(\ba) = \mu(S_k)^{-1} \int_{S_k} \om(g\ba)d\mu(g), 
\] 
where $d\mu(g)$ is the invariant measure on $E$ and $g \in E$. It can be 
shown~\cite{haag93} that the sequence $\om_k$ converges to a positive linear 
form $\om_E$, which is invariant under $E$. Using this form we obtain a 
representation of $\Aob$ on a state space $\mathcal{H}_E$ carrying a unitary 
representation of $E$. Equivalently, we could have used this procedure to 
average the background state $\Q$, to obtain the desired representation.

\section{Conclusions and future directions} 
 
In this work, we have presented two main results. The first of these is a 
general procedure for construction of new Hilbert spaces for loop quantum 
gravity for non-compact spaces. We used an analogy with thermal field theory to 
construct a non-standard representation of the classical algebra of 
observables. A key ingredient for this construction was the use of a background 
state which is analogous to the thermal ground state. 
 
We also presented a possible candidate which can be used as a background state. 
This is a weave state in that it approximates the classical flat Euclidean 
metric on $\mathbb{R}^3$. The advantage of this state over previous weave 
constructions is that it is not an eigenfunction of the three-geometry. Rather, 
it is peaked both in the connection and the spin-network pictures. This was our 
second main result. 
 
We would like to conclude with a discussion of some open issues and future 
directions. 
\begin{enumerate} 
\item The properties of the state $\Q$ deserve to be better studied. In 
particular, we would like to investigate any possible relation between $\Q$ and 
coherent states which may be defined on the group SU(2). 
 
\item We would also like to study the low-energy sector of the Hilbert 
space constructed with $\Q$ as a background state and look at its 
relation to the standard Fock space of gravitons. In this context, it 
would also be interesting to understand the connection of our work 
to~\cite{iwasaki93,iwasaki94}. 
 
\item We are in the process of computing the spectra of the area and volume 
operators in the new Hilbert space defined by $\Q$ to verify the intuitive 
picture of areas and volumes fluctuating around flat space values. 
 
\item A quantum positivity of energy theorem was proved in~\cite{thiemann98}. 
However, as we have shown, the Hilbert space on which that result was proved is 
not applicable to the study of non-compact spatial geometries. We believe that 
our construction provides the proper arena for questions of this nature and are 
currently investigating the properties of suitably defined ADM energy and 
momentum operators on our Hilbert space. 
\end{enumerate} 
 
This work is a step in the direction of making contact between the 
non-perturbative quantisation of gravity and the picture of graviton physics 
which arises from standard perturbative quantum field theory. A lot more work 
needs to be done before the relation between the two is completely clarified. 
 
\section*{Acknowledgements} 
 
We would like to thank Chris Isham, Carlo Rovelli and Lee Smolin for 
discussions and motivation. We would also like to thank Abhay Ashtekar 
for his comments on an earlier version of this manuscript. S.G.\ was 
supported in part by NSF grant PHY-9514240 to the Pennsylvania State 
University and a gift from the Jesse Phillips Foundation. 
 
\bibliography{references} 
 
\end{document}